\documentclass[twocolumn,showpacs,preprintnumbers,amsmath,amssymb]{revtex4}
\usepackage{dcolumn}
\usepackage{bm}
\usepackage{graphicx}
\usepackage{color}
  \def\epspdffile#1{\scalebox{0.25}{\includegraphics{#1.eps}}}
\def\rational#1#2{{\mathchoice{\textstyle{#1\over#2}}%
  {\scriptstyle{#1\over#2}}{\scriptscriptstyle{#1\over#2}}{#1/#2}}}
\def\half{\rational12}		    
\def\quarter{\rational14}	    
\def\rmsub#1#2{#1_{\mbox{\tiny #2}}} 
\def\rmsup#1#2{#1^{\mbox{\tiny #2}}} 
\def\rmss#1#2#3{#1_{\mbox{\tiny #2}}^{\mbox{\tiny #3}}}	
\def\defn{\equiv}		    
\def\det{\mathop{\rm det}}
\def\tr{\mathop{\rm tr}}
\def\ln{\mathop{\rm ln}}
\def\asqtad{{\sc ASQTAD}}	    
\def\nth{\rmsup{n}{th}}		    
\def\dt{\delta\tau}		    
\def\dH{\delta H}		    
\def\md{MD}			    
\def\mc{MC}			    
\def\hmc{HMC}			    
\def\rhmc{RHMC}			    
\def\pacc{\rmsub{P}{acc}}	    
\def\tD{\mbox{D}\kern-0.65em\raise0.15ex\hbox{/}\kern0.15em} 
\def\sD{\mbox{\scriptsize D}\kern-0.5em\raise0.15ex\hbox{\scriptsize/}}
\def\ssD{\mbox{\tiny D}\kern-0.42em\raise0.15ex\hbox{\tiny/}}
\def\Dslash{{\mathchoice{\tD}{\tD}{\sD}{\ssD}}}
\def\dslash{\hbox{\(\partial\)}\kern-0.5em\raise0.15ex\hbox{/}} 
\def\m{{\cal M}}		    
\def\Ml{\rmsub{\m}{\(\ell\)}}	    
\def\Ms{\rmsub{\m}{s}}		    
\def\ml{\rmsub{m}{\(\ell\)}}	    
\def\ms{\rmsub{m}{s}}		    
\def\phil{\rmsub{\phi}{\(\ell\)}}   
\def\phis{\rmsub{\phi}{s}}	    
\def\nmd{\rmsub{n}{MD}}		    
\def\nmc{\rmsub{n}{MC}}		    
\def\res{\mathop{\rm res}}	    
\def\resmd{\rmsub{\res}{MD}}	    
\def\resmc{\rmsub{\res}{MC}}	    
\def\rl{\rmsub{r}{\(\ell\)}}	    
\def\rs{\rmsub{r}{s}}		    
\def\rmcl{\rmss{r}{\(\ell\)}{MC}}   
\def\rmcs{\rmss{r}{s}{MC}}	    
\def\rmdl{\rmss{r}{\(\ell\)}{MD}}   
\def\rmds{\rmss{r}{s}{MD}}	    

\def\Ncg{\rmsub{N}{cg}}
\def\Nflops{\rmsub{N}{flops}}

\def\Sg{\rmsub{S}{G}}		    
\def\Sf{\rmsub{S}{F}}		    
\def\Sfmd{\rmss{S}{F}{MD}}
\def\naive{na\"{\i}ve}
\begin{document}

\preprint{APS/123-QED}

\title{Accelerating Staggered Fermion Dynamics with the\\Rational Hybrid
Monte Carlo (RHMC) Algorithm}

\author{M. A. Clark}
  \email{mikec@bu.edu}
  \affiliation{%
    Center for Computational Sciences,  Boston University,\\
    3 Cummington Street, Boston,  MA 02215,  United States of America}%
\author{A. D. Kennedy}
  \email{adk@ph.ed.ac.uk}
  \affiliation{
    School of Physics, The University of Edinburgh,\\
    Mayfield Road, Edinburgh, EH9 3JZ, United Kingdom}%
\date{11 September 2006}
\begin{abstract}
  \noindent Improved staggered fermion formulations are a popular choice for
  lattice QCD calculations.  Historically, the algorithm used for such
  calculations has been the inexact R algorithm, which has systematic errors
  that only vanish as the square of the integration step-size.  We describe how
  the exact Rational Hybrid Monte Carlo (RHMC) algorithm may be used in this
  context, and show that for parameters corresponding to current
  state-of-the-art computations it leads to a factor of approximately seven
  decrease in cost as well as having no step-size errors.{\pretolerance=10000
  \parfillskip=0pt\par}
\end{abstract}
\maketitle

Staggered fermions are a popular choice for lattice QCD calculations: they are
generally thought of as being much cheaper than other fermion formulations, and
have a remnant of the continuum theory's chiral symmetry protecting them from
additive mass renormalization.  In order to overcome the infamous fermion
doubling problem, the spin degrees of freedom are spatially separated reducing
the \(2^4\) fermion ``tastes'' to four.  To further reduce these four tastes to
a single quark (or doublet) the fourth (square) root of this kernel is taken.
While this is a valid prescription in the continuum limit, it is unclear what
unwanted side-effects this introduces to the lattice theory.  At best this
treatment is an algorithmic nuisance, at worst it renders the theory non-local,
non-universal, and non-unitary rendering any calculations made using such a
formulation questionable.

The reduction from 16 to 4 fermions also introduces \(O(a^2)\) unphysical
taste-mixing interactions, where \(a\) is the lattice spacing.  There are
various improved formulations of staggered fermions, all of which redefine the
Dirac operator by some local averaging of the gauge field variables; the paths
and coefficients of this averaging are chosen to minimize these unphysical
interactions.  The most popular formulation is the \asqtad\ prescription, where
the Dirac operator is constructed from a gauge covariant derivative consisting
of the sum of the product of 1, 3, 5 and 7 link variables
\cite{Orginos:1999cr}.

After integrating out the Grassman-valued quark fields, the 2+1 quark flavor
QCD partition function is given by a functional integral over gauge fields,
\vspace{-2mm}
\begin{equation*}
  Z = \int dU \det(\Ml)^\half \det(\Ms)^\quarter e^{-\Sg},
\vspace{-2mm}
\end{equation*}
where \(\Sg\) is the gauge action (for \asqtad\ fermions this is usually the
one-loop improved Symanzik action) and \(\m = -\Dslash^2 + 4m^2\) is the
fermion matrix, with \(\Dslash\) being the staggered covariant derivative and
\(m\) the quark mass.  The determinants represent the (\(\ell\))ight and
(s)trange quark contributions to the vacuum.  It is the matrix square root and
fourth root that make the formulation so algorithmically problematic: the
Hybrid Monte Carlo (\hmc) \cite{Duane:1987de} algorithm commonly used for
lattice calculations with dynamical fermions not being directly applicable.

The algorithm commonly used for performing these calculations is the Hybrid~R
algorithm \cite{Gottlieb:1987mq}.  First the identity \(\det\m = \exp\tr\ln\m\)
is used to write the action as
\vspace{-2mm}
\begin{equation*}
  S = \Sg - \half \tr\ln \Ml - \quarter \tr\ln \Ms,
\vspace{-2mm}
\end{equation*}
then a Gaussian-distributed ``fictitious'' momentum field \(\pi\), taking
values in the Lie algebra \(su(3)\) on links, is introduced so a Hamiltonian
\(H = \half \pi^2 + S\) may be defined.  Configurations \((\pi,U)\) are then
generated with probability proportional to \(e^{-H}\) by alternating two Markov
steps, refreshment of the momenta from a Gaussian heatbath and evolution of the
gauge fields following the \md\ (\md) trajectory defined by this Hamiltonian
for time \(\tau\).  The \md\ trajectory is approximated by means of a numerical
integrator with a step-size of \(\dt\), so these steps give an ergodic Markov
process with the desired fixed point up to step-size errors.  Usually the
second order leapfrog (2LF) integrator is used which requires a single force
evaluation per step, and this leads to errors of \(O(\dt^2)\) in the
coefficients in the action.

The fermionic contribution to the force acting on the gauge fields is
approximated by means of a noisy estimator for each trace appearing in the
action: this requires the introduction of an auxiliary ``noise'' field that is
refreshed after every update.  Each fermion force evaluation requires that the
inverse fermion matrix be applied to its auxiliary field, typically by means of
a conjugate gradient (CG) solver.  The cost of the algorithm increases
dramatically as the fermion mass is taken to zero, and the cost of the light
quarks contribution dominates that of the strange.

Additional \(O(\dt)\) errors are introduced by using noisy estimates of the
fermionic force, regardless of the numerical integrator used.  These may be
made to vanish by refreshing the auxiliary field at an asymmetric time within
each update step, dependent on the number of fermions; doing so leads to an
integrator that is neither area-preserving nor reversible, and therefore the
step-size errors cannot be removed by means of a Metropolis acceptance test as
with HMC.  The R algorithm is thus necessarily an inexact algorithm with
\(O(\dt^2)\) errors.

An alternative approach is to represent each fermionic determinant as a
Gaussian integral over a bosonic ``pseudofermion'' field \(\phi\): the
resulting fermion action is \(\Sf = \phil^\dagger \Ml^{-\half} \phil +
\phis^\dagger \Ms^{-\quarter} \phis\).  At this point the conventional \hmc\
algorithm fails as one can neither evaluate the \(\nth\) root of a matrix
directly nor calculate its derivative with respect to the gauge field as
required for the \md\ evolution of the system.  However, it is possible to
replace the matrix kernel with an approximation that can be evaluated and
differentiated.  There are two obvious approximations: polynomial
\cite{deForcrand:1996ck,Frezzotti:1997ym} and rational \cite{Kennedy:1998cu,
Clark:2004cp}.  Polynomial approximations do not require the explicit
evaluation of a matrix inverse, but they typically have to be of much higher
degree than the corresponding rational approximations with the same error: for
example, a polynomial approximation to \(1/\m\) for an hermitian matrix \(\m\)
is equivalent to computing the inverse by means of a Jacobi iteration scheme,
and this is well-known to be far inferior to other iterative solvers such as
CG.  With this in mind, we focus on optimal rational approximations (generally
found using the Remez algorithm).  We replace the square and fourth root
kernels by rational approximations, and continue as if we would for \hmc, for
which each complete update consists of a sequence of the following Markov steps
\begin{itemize}
    \vspace{-1mm}
\item Momentum refreshment heatbath using Gaussian noise
  (\(P(\pi)\propto e^{-\pi^2/2}\)).
    \vspace{-1mm}
\item Pseudofermion refreshment (\(P(\phi)\propto \m^\frac1{2\alpha}\eta\),
  where \(P(\eta)\propto e^{-\eta^2/2}\), \(\alpha=2\)(4) for light (strange)).
    \vspace{-1mm}
\item MDMC update, built out of
  \begin{itemize}
    \vspace{-1mm}
  \item A (\md) trajectory consisting of \(\tau/\dt\) steps.
    \vspace{-1mm}
  \item A Metropolis accept/reject with probability \(\pacc=\min(1,e^{-\dH})\).
  \end{itemize}
\end{itemize}
The key advantage over the R algorithm is that the Metropolis test at the end
of the trajectory stochastically corrects any finite step-size errors
introduced through the discretized \md\ integration.

It is necessary to generate the pseudofermion fields at the start of each
trajectory from a heatbath.  For the functions of interest, the roots and poles
of optimal (in the minimax \(= L_\infty = \) Chebyshev norm) rational
approximations are real and non-degenerate: this motivates a fermionic action
\(\Sf = \phil^\dagger [\rmcl(\Ml)]^2 \phil + \phis^\dagger [\rmcs(\Ms)]^2
\phis\), where \(\rmcl(x)\approx x^{-\quarter}\) and \(\rmcs(x)\approx
x^{-\rational18}\) are valid over the spectral bounds of the operator.  These
rational approximations must in some sense be exact: typically this means the
maximum error is either less than the tolerance used in the solver or, more
conservatively, less than the unit of least precision of double precision
arithmetic (\(\approx 10^{-15}\)).

When written in partial fraction form (we consider only the case where the
numerator and denominator polynomial degrees are equal) \(r(x) = \alpha_0 +
\sum_{k=1}^{n} \frac{\alpha_k}{x+\beta_k},\) where \(n\) is the degree of the
approximation.  It transpires that for the cases of interest all the
\(\alpha_k\) and \(\beta_k\) are real and positive: this ensures numerical
stability of the method.

A multi-shift solver \cite{Frommer:1995ik} can be used to evaluate the rational
approximation with a cost similar to that of the inversion of the original
matrix \(\m\), thus both the pseudofermion heatbath and the evaluation of the
action for the Metropolis step can be carried out within a single Krylov space.
A difficulty arises when the fermionic force has to be computed, as this seems
to depend upon the derivative of the square of the rational approximation
\(r\), which involves double poles; this is obviously undesirable, as \naive ly
it doubles the cost as compared to the R algorithm.  However, there is no
reason why the \md\ and \mc\ approximations need be the same: one is free to
choose the \md\ approximation to avoid the square, and also use a lower degree
approximation should one so wish.  Indeed, the optimal approximation to
\(\m^{2\alpha}\) is not the square of that for \(\m^\alpha\) anyhow.  The \md\
action is thus written
\vspace{-1mm}
\begin{equation}
  \Sfmd = \phil^\dagger \rmdl(\Ml) \phil + \phis^\dagger \rmds(\Ms) \phis,
  \label{eq:md-action}
\vspace{-1mm}
\end{equation}
where the approximations \(\rmdl(x) \approx x^{-\half}\) and \(\rmds(x) \approx
x^{-\quarter}\) are again valid over the appropriate spectral ranges.

The derivative of each of the bilinear terms that appear in equation
(\ref{eq:md-action}) is written as
\vspace{-2mm}
\begin{eqnarray}
  \Sf' & = & - \sum_{k=1}^{n} {X^{(k)}}^\dagger \m' X^{(k)}
    \label{eq:force-expense} \\
    & = & - \tr\Bigl[\m' \sum_{k=1}^{n} X^{(k)}\otimes{X^{(k)}}^\dagger\Bigr],
      \label{eq:force-cheap}
\vspace{-2mm}
\end{eqnarray}
where \(X_k = \sqrt{\alpha_k} (\m+\beta_k)^{-1}\phi\).  The calculation of
\(\m'\) must be done explicitly using the chain rule and represents a
considerable computational expense because of the complicated nature of the
\asqtad\ operator.  The only dependence on the pseudofermion fields is in
\(X^{(k)}\), and even if there are many such fields one need only calculate
\(\m'\) once.  For \(n>3\) it is more expensive to use matrix-vector operations
(\ref{eq:force-expense}) than matrix-matrix operations (\ref{eq:force-cheap}),
hence we use the latter~\cite{Clark:2004cp}.

The \naive\ cost of \rhmc\ is similar to that of the R algorithm at the same
step-size: however, the actual cost is determined by the integration step-size
that can be used, and the algorithms' respective autocorrelations.

Before we present results from such a comparison, we describe a reformulation
of the fermion action which greatly reduces the cost of \rhmc.  It has been
shown that ``mass preconditioning'' the fermion force leads to a large
reduction in the computational cost \cite{Hasenbusch:2001ne}; the fermion
kernel is multiplied by the inverse of a similar kernel with a larger mass, and
the desired distribution is then recovered by also introducing a new
pseudofermion field with this larger mass kernel.  This is beneficial because
the resulting pseudofermion kernels are less singular, and thus lead to smaller
forces acting on the gauge fields; this in turn permits the use of a larger
integration step-size before the symplectic integrator becomes unstable.
A multiple timescale numerical integrator where the mass preconditioned light
pseudofermion force is evaluated less often than that of the new pseudofermion
then gives further speed-up: when tuning such an algorithm the product of each
step-size with the magnitude of the corresponding force contribution should be
the same to a first approximation.

\begin{figure}
  \epspdffile{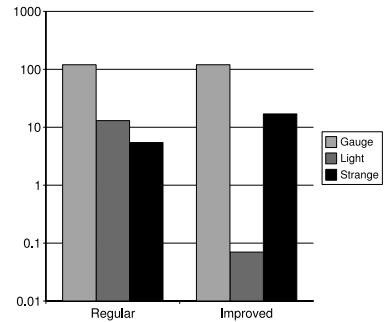}
  \caption{\label{fig:force} A comparison of the relative \(L_\infty\)
    magnitude of the forces of the original \(2+1\) \rhmc\ \asqtad\
    formulation, and that of the mass preconditioned one.  In the former the
    light quark force is more than three times that of the strange
    contribution, in the latter it is two orders of magnitude smaller
    (\(V=4^4\), \(\beta=6.76\), \(\ml=0.01\), \(\ms=0.05\)).}
\end{figure}

In previous work \cite{Urbach:2005ji} mass preconditioning was applied to \(2\)
flavor Wilson fermion calculations, but neither to \(2+1\) quark flavors nor to
staggered fermions, both of which lend themselves particularly well to this
technique because the strange quark can be used as a preconditioner, as shall
now be explained.  The \(2+1\) flavor staggered determinant may be written as
\vspace{-2mm}
\begin{equation*}
  \det(\Ml)^\half\det(\Ms)^\quarter =
  \left(\frac{\det(\Ml)}{\det(\Ms)}\right)^\half \, \det(\Ms)^\rational34,
\vspace{-2mm}
\end{equation*}
where the light quark has been ``mass preconditioned'' by the strange quark.
The corresponding action is
\vspace{-2mm}
\begin{eqnarray}
  \Sf &=& \phil^\dagger \left(\frac{\Ms}{\Ml}\right)^\half\phil +
    \phis^\dagger \Ms^{-\rational34} \phis \nonumber \\
  &=& \phil^\dagger \left(\frac{\Ml + \delta m^2}{\Ml}\right)^\half\phil +
    \phis^\dagger \Ms^{-\rational34} \phis \nonumber \\
  &\approx& \phil^\dagger \rl(\Ml) \phil + \phis^\dagger \rs(\Ms) \phis,
    \label{eq:action-prerat} 
\vspace{-2mm}
\end{eqnarray}
where in the second line use has been made of the fact that the staggered quark
mass is additive, \(\Ms = \Ml + \delta m^2\) with \(\delta m^2 \defn 4(\ms^2 -
\ml^2)\), so that the first term may be written in terms of a single rational
function of the light quark kernel in equation~(\ref{eq:action-prerat}).  The
cost of evaluating the action is almost unchanged since no extra fields have
been introduced, rather better use is being made of the already present strange
quark.  Since the light quark is preconditioned by the strange quark we expect
that its contribution to the total force will be small in comparison to the
``triple strange''.  This is advantageous because the triple strange force is
much cheaper to evaluate since it is a rational approximation of the more
massive strange quark kernel.  Thus the hierarchy of the integrators is to have
the gauge field force updates (which are cheap compared with fermionic force
evaluations) on the smallest time scale, the triple strange on an intermediate
scale, and the preconditioned light quark on the largest scale.  We note that
choosing the strange quark to precondition the light quarks is merely down to
the former's presence: if the strange quark were not already present then the
optimal precondittioned would certainly have a difference mass.
Figure~\ref{fig:force} is an example of the effect the mass conditioning has on
the forces: the vast reduction in magnitude of the light quark contribution is
evident.
A nice side effect of such mass preconditioning is that the approximation
degree required for the preconditioned light kernel is much less than for the
original one because its kernel is much better conditioned.

As well as mass preconditioning, there are other improvements that can be made
to the \rhmc\ algorithm.  The \(\nth\) root multiple pseudofermion trick
\cite{Clark:2004cq} can be used to allow a larger integration step-size at
constant acceptance rate: here one pseudofermion with kernel \(\m^\alpha\) is
replaced by \(n\) pseudofermions each with kernel \(\m^{\alpha/n}\).  This
method removes the integrator instability, after which it becomes advantageous
to use higher-order or improved integrators~\cite{Clark:2006fx}, e.g., a second
order minimum norm (2MN) integrator~\cite{Takaishi:2005tz}.  The tolerance of
the \md\ solver can be tuned on a per-shift basis; in particular the tolerances
of least well-conditioned shifts can be loosened considerably with little
affect on the acceptance rate.  This leads to a large reduction in the number
of CG iterations~\cite{Clark:2005sq}.

In order to test the improved \rhmc\ algorithm, a suitably challenging set of
parameters were chosen, as specified in Table~\ref{table-results}.  These
parameters also represent a recent run of the R algorithm performed by the MILC
collaboration, so a direct comparison is possible.

\begin{table}[htb]
  \advance\tabcolsep by2pt
  \begin{tabular}{lcc}
    \hline\hline\noalign{\vskip2pt}
    & R \cite{MILC} & \rhmc \\[1pt] \hline\noalign{\vskip1pt}
    \(\tau/\dt\) & \(300\) 
      & \(\rmsub{5}{s,\(\ell\)}, \rmsub{25}{gauge}\) \\ [1pt]
    \((\nmd,\nmc)\) & ---
      & \(\rmsub{(6,10)}{s}\), \(\rmsub{(4,7)}{\(\ell\)}\) \\[1pt]
    \(\resmd\) 
      & \(\rmsub{(5\cdot10^{-5})}{s,\(\ell\)}\) 
      & \(\rmsub{(10^{-4})}{s}\), \(\rmsub{(10^{-5})}{\(\ell\)}\) \\[1pt]
    \(\resmc\) & --- & \(\rmsub{(10^{-10})}{s,\(\ell\)}\) \\[1pt] 
    \(\Ncg\) & 344,988 & 70,230 \\ [1pt]
    \(\Nflops\) & \(5.96\times 10^{14}\) & \(8.54\times 10^{13}\)\\ [1pt]
    \(\langle\pacc\rangle\) & --- & 0.775(23) \\[1pt]
    \hline\hline
  \end{tabular}
  \caption{Parameters and performance: \(n\) denotes the rational approximation
    degree, \(\pacc\) the Metropolis acceptance rate, and \(\res\) the CG
    solver residual. The subscripts s and \(\ell\) indicate the fermion
    flavor. \(\Ncg\) and \(\Nflops\) are the total CG iterations and floating
    point operations per trajectory respectively.  The computations were
    performed on a \(V=24^3.64\) lattice with \(\beta=6.76\), \(\ml=0.005\),
    \(\ms=0.05\), and \(\tau=1.0\).}
  \label{table-results}
\end{table}

From previous studies of the R algorithm it has been found empirically that
choosing an integrator step-size \(\dt\approx\rational23\ml\) leads to
step-size errors that are deemed to be negligible.  On the other hand, deciding
on the optimum step-sizes for multiple timescale \rhmc\ algorithm is a more
involved process.  When constructing the multi-scale integrator, it was decided
to evaluate the gauge force using a fourth order integrator
\cite{Takaishi:2005tz} to minimize this part of the cost.  After constructing
the multiple timescale integrator, a single \md\ trajectory was used to measure
the force contributions in order to estimate the step-size ratios.  It was
observed that the cost of the calculation was dominated by the triple strange;
this in turn was dominated by the calculation of the matrix derivative, and not
the matrix inversion (this effect is exacerbated by the tolerance loosening
optimization); thus any further mass preconditioning would be of little
benefit.  To further reduce the algorithm cost, the \(\nth\) root multiple
pseudofermion formulation was used for the triple strange, with \(3\)
pseudofermions, this allowed for a large increase of step-size, so much so in
fact that the optimal parameters were found to be those where the light and
triple strange were all on the same time scale again, i.e., a two level
integrator separating the fermion and gauge timescales.  It is interesting to
note that the optimal algorithm for 2+1 flavour \asqtad\ fermions requires both
mass preconditioning and the \(\nth\) root trick, emphasizing the opinion
expressed in \cite{Clark:2004cq} that these approcahes can be complimentary.

A summary of the optimal parameters and costs is given in
Table~\ref{table-results}.  Perhaps the most interesting result is the factor
of 7 reduction in the number floating point operations per trajectory relative
to the R algorithm.  It must be emphasized this factor does not take into
\begin{figure}[ht]
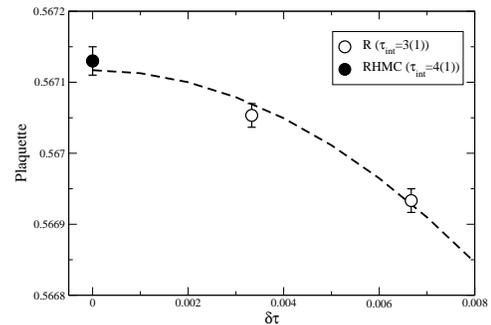

  \epspdffile{plaq}
  \caption{\label{fig:plaquette} An extrapolation of the plaquette in \(\dt\)
    comparing the exact RHMC result with the inexact R result~\cite{MILC}.}
\end{figure}
account the additional benefit of using an exact algorithm over an inexact one,
illustrated in figure~\ref{fig:plaquette}.  The autocorrelation lengths of the
plaquette using the algorithms are similar, indeed they are essentially the
same after dividing by \rhmc's acceptance rate, and this is expected to be true
for all other autocorrelations too.

In this letter, we have described the exact \rhmc\ algorithm, and how this
algorithm can be used to greatly accelerate the generation of lattice QCD gauge
field ensembles with dynamical \asqtad\ fermions compared to current techniques
using the inexact R algorithm.  Through a switch of algorithm, a production job
that takes a year becomes at worst a matter of months, while at the same time
removing a source of systematic error.

\section*{Acknowledgments}
\vspace{-3mm}
We wish to thank Steve Gottlieb for providing us with a thermalized lattice.

The development and computer equipment used in this calculation were funded by
the U.S. DOE grant DE-FG02-92ER40699, PPARC JIF grant PPA/J/S/1998/00756 and by
RIKEN. This work was supported by PPARC grants PPA/G/O/2002/00465,
PPA/G/S/\-2002/00467 and PP/D000211/1.


\end{document}